\documentclass{article}
\usepackage{amsmath,amsfonts,amssymb}
\usepackage{tensor}
\usepackage{natbib}
\usepackage{iclr2016_conference,times}
\usepackage{hyperref}
\usepackage{url}
\usepackage{todonotes}
\usepackage{xspace}
\usepackage{vector}

\newcommand{\weit}{Weitzenb\"{o}ck\spacer}

\newcommand{\spacer}{\@\xspace}

\newcommand{\chris}[3]{\ensuremath \left\{ ^{#1}_{#2#3} \right\}}

\title{Principal Autoparallel Analysis: \\ Data Analysis in Weitzenb\"{o}ck Space}
 \author{Stephen Marsland \\
 School of Engineering and Advanced Technology\\ Massey University\\ Palmerston North, New Zealand \\ \url{s.r.marsland@massey.ac.nz}
 \And 
 Carole Twining \\
 Imaging Science and Biomedical Engineering \\ The University of Manchester\\ Manchester, UK \\ \url{carole.twining@manchester.ac.uk}
 }

\begin{document}
\maketitle

\begin{abstract}
The statistical analysis of data lying on a differentiable, locally Euclidean, manifold introduces a variety of challenges because the analogous measures to standard Euclidean statistics are local, that is only defined within a neighbourhood of each datapoint. This is because the curvature of the space means that the connection of Riemannian geometry is path dependent. In this paper we transfer the problem to \weit space, which has torsion, but not curvature, meaning that parallel transport is path independent, and rather than considering geodesics, it is natural to consider autoparallels, which are `straight' in the sense that they follow the local basis vectors. We demonstrate how to generate these autoparallels in a data-driven fashion, and show that the resulting representation of the data is a useful space in which to perform further analysis.
\end{abstract}

\section{Introduction}

There has been a lot of recent interest in the analysis of data lying on a differentiable, locally Euclidean, manifold, rather than in standard Euclidean space. Such data can arise in lots of application areas, but the analysis of shape has been a particular interest~\citep{Pennec,Fletcher,Sommer}. The particular challenge of statistics on manifolds is that the space is not uniformly flat, meaning that while many of the standard techniques of Euclidean data analysis have analogues, they may not be uniquely defined, and they are {\em local}, that is only defined within an infinitesimal neighbourhood of each datapoint. 

This last point is crucial. The particular differentiable manifold that is considered is the Riemannian manifold and the `tools' of data analysis on such a manifold  (the Riemannian metric, and the connection) are built at the scale of infinitesimal neighbourhoods, through the computation of tangent vectors and so tangent spaces at each point of the manifold. The metric compares pairs of tangent vectors defined at the same point, defining the inner product between them, which means that the length of vectors can be computed, and integrated to compute distances along paths in the manifold. Then, in order to compare the tangent spaces at different points on the manifold, a connection is required that describes how to transport vectors defined at one point to another. 

The curvature of a manifold means that parallel transport is not path-independent, so that transporting the same vector between two places by different paths will generally produce different vectors at the endpoint. For example, consider the case of parallel transporting a vector on the shell of a sphere in normal three dimensional space, such as (approximately) the surface of the earth. We start at the north pole, move down to the equator, around the equator for one quarter of the earth's circumference, and then back to the north pole. Comparing the resulting vector with one that did not move at all, we can see that the transported one is now at $90^{\circ}$ to the original version. Importantly for data analysis, the lack of path independence means that it is impossible to define a data basis and thus a global coordinate system on the manifold. 

In addition to curvature, there is another geometric invariant of differentiable manifolds, which is torsion. Far less commonly considered,  torsion of a manifold means that parallel transporting around an infinitesimal parallelogram does not return exactly to the original point, so that loops are not closed; the additional translation required to close a loop gives a measure of the torsion of that region. The reason why torsion is not used is that there exists a particular choice of connection, the Levi-Civita connection, that has curvature but everywhere zero torsion, thus simplifying things. In this paper we will consider a differentiable, locally Euclidean, manifold endowed not with the Levi-Civita connection, but with the Weitzenb\"{o}ck connection, which has torsion, but where the curvature is uniformly zero. 

The Weitzenb\"{o}ck connection can be defined in terms of Cartan's moving frames~\citep{Cartan1922}, which could be used explicitly as a basis for data if they were known. Unfortunately, they are data dependent, and the point of this paper is to derive an algorithm that constructs an estimate of the \weit frame at each point of the manifold. We introduce the idea of working on data spaces endowed with the Weitzenb\"{o}ck connection, and derive an algorithm that generates autoparallels in a data-driven fashion on the dataspace, enabling the mapping of the data manifold. An autoparallel is the `straightest' path on the manifold according to the directions of the moving frame at each point. Our algorithm is, of necessity, iterative, because we can only make local estimates of the moving frame, and hence the autoparallels, and need to remove each principal autoparallel to find the next most common direction in the data. 

\weit space has also been used to describe an autoparallel version of General Relativity~\citep{Hammond}, the shears that sometimes occur in crystals, when points where the regular structure of the atoms of the crystal change~\citep{Kleinert}, and non-holonomic mechanics~\citep{Fernandez}.

\section{Mathematical Formulation}\label{sec:mathform}

We here present a brief overview of the concepts required to understand the geometry of Weitzenb\"{o}ck space. For further details, see for example~\citep{Weit,essay,Guo}; for an alternative description of torsion based on the soldering form, see~\citet{Michor}.

\subsection{Torsion}

Given a differentiable, locally Euclidean, manifold with an affine connection, torsion can be defined  as the amount by which parallel transport parallelograms do not close. Consider starting with a vector $\Delta x$ at a point $x$, and transporting it in a direction $x^1$, followed by another direction $x^2$, and also by transporting it in direction $x^2$ followed by $x^1$. In a Riemannian manifold with curvature the parallelogram formed by following both paths is closed by construction, but the resultant vector will not point in the same direction if the space is curved. In a manifold with torsion, but not curvature, the parallelogram will not be closed, but the resultant vectors will point in the same direction; see Figure 1 of~\citet{essay} for a picture of this. The general case of a differentiable manifold with both curvature and torsion is known as a Riemann-Cartan manifold. 

In terms of the covariant derivatives of affine connection $\nabla$ of vector fields $X$ and $Y$ the torsion can be defined as (where $[X,Y]$ is the Lie bracket of $X$ and $Y$):
\begin{equation}
T(X,Y) =  \nabla_{X}Y - \nabla_{Y}X -[X,Y].
\end{equation}
 
In a coordinate basis, the Lie Bracket of the basis vectors
vanishes by definition, and so the torsion reduces to (where $\Gamma\indices{^\mu_{\alpha\beta}}$ are the connection in component form):
\begin{equation}
T\indices{^\mu_{\alpha\beta}} = \Gamma\indices{^\mu_{\alpha\beta}}-\Gamma\indices{^\mu_{\beta\alpha}},
\end{equation}
which is the anti-symmetric part of the connection in the
coordinate basis. 


\subsection{The Weitzenb\"{o}ck Frame}

Consider standard Euclidean space $\mathbb{R}^n$, which we associate with a Weitzenb\"{o}ck space $\mathbb{W}$. 
Since we are working in Euclidean space, there is a natural coordinate frame, which is global, each direction of which corresponds to a unit-length tangent vector with only one non-zero coordinate component. 

At any point $x \in \mathbb{W}$ we can define a local basis using $n$ orthonormal basis vectors $\{ e_a(x), a = 1 \ldots n \}$, with coordinate components that depend upon their position (where there are $N$ datapoints):
\begin{equation}
e_{a}(x) = \{e\indices{_{a}^{\mu}}(x): \mu = 1,\ldots N\}.
\end{equation}

These basis vector fields are the position-dependent Weitzenb\"{o}ck frame, a particular example of a moving frame~\citep{Cartan1922}, also known as a {\em vielbein}, or specifically in 2D, a {\em zweibein}, and 3D a {\em dreibein}). The frame is orthonormal at every point by construction, but anholonomic, since the Lie derivative does not vanish. This means that the order in which the basis vectors are followed matters: consider starting at a point $\mathbf{x}$ on the manifold and following the two paths given by going in direction $x^{1}$ then direction $x^{2}$, and vice versa:
\begin{eqnarray}
&&x(0) \rightarrow x(0) + \Delta x^{1} \rightarrow \left(x(0) + \Delta x^{1}\right) + \Delta x^{2}, \nonumber \\
&&x(0) \rightarrow x(0) + \Delta x^{2} \rightarrow \left(x(0) + \Delta x^{2}\right) + \Delta x^{1}. 
\end{eqnarray}

For an arbitrary vector $Y$ there is a simple relationship between the components in the coordinate frame and Weitzenb\"{o}ck frames:
\begin{equation}
Y^{\mu}(x) = e\indices{_a^\mu}(x)Y^{a}(x),
\end{equation}
 
We can also define torsion in this frame basis:
\begin{equation}
T\indices{^a_{bc}} = -[e_{b},e_{c}]^{a}.
\end{equation}

%

A vector field $X$ is everywhere parallel if and only if in the \weit frame:
\begin{equation}\label{eq:paradef} 
X^{a}(x) \equiv f(x) X^{a}(y) \:\: \forall \:\: x, y \in \mathbb{W} \:\: \& \:\:
\forall \:\: a = 1,\ldots n,
\end{equation}

\noindent where $y$ is a reference point and $f(x)$ is a scalar function.

We can now start to consider how a tangent vector $\Delta x$ at $x \in \mathbb{W}$ describes an infinitesimal change in the position of that point. By using the Euclidean metric on $\mathbb{R}^n$ space we induce a metric on $\mathbb{W}$:
\begin{equation}\label{eq:gdef}
|\Delta x|^{2} \equiv
g_{\mu\nu}(x)\Delta x^{\mu}\Delta x^{\nu} = g_{\mu\nu}(x) e_a^{\mu}(x) e_b^{\nu} (x)  = \delta_{ab}, 
\end{equation}
\noindent where $\delta_{ab}$ is the Kronecker delta.

The change in coordinate components of any arbitrary vector as it moves from $x$ to $x + \Delta x$ under parallel transport in a differentiable manifold is described via the connection (although note that with the addition of torsion, the connection does not retain the index-symmetry of the Christoffel symbols familiar to Riemannian geometry with the Levi-Civita connection):
\begin{equation}\label{eq:connectdef}
\delta^{\|}Y^{\mu}(x) \doteq - \Gamma\indices{^\mu_{\alpha\beta}}(x)Y^{\alpha}(x)\Delta x^{\beta}.
\end{equation}

Combining (\ref{eq:gdef}) and (\ref{eq:connectdef}) we see that we can define an everywhere-parallel vector field $X$ by taking the value of the field at a single point, and then parallel transporting it to every point. This gives the relation between the \weit frame $\{e_{a}\}$ and the \weit connection $\Gamma$ defined by that frame:
\begin{equation}
\partial_{\beta}e\indices{_a^\mu}(x) \equiv - \Gamma\indices{^\mu_{\alpha\beta}}e\indices{_a^\alpha}(x), \:\: \partial_{\alpha} \doteq \frac{\partial}{\partial x^{\alpha}}.
\end{equation}

Clearly, the parallel transport is independent of the path taken, and hence the curvature is zero everywhere by construction, and the space possesses absolute parallelism. To see that the space does indeed have torsion, consider the construction of an infinitesimal parallelogram using the frame basis vectors:
\begin{equation}
 e\indices{_{a}^\mu}(x) +  e\indices{_{b}^\mu}(x + \varepsilon e_{a}(x))
 \ne  e\indices{_{b}^\mu}(x) +  e\indices{_{a}^\mu}(x + \varepsilon e_{b}(x)), \varepsilon \ll 1,
\end{equation}
which means the (infinitesimal) parallelogram will not close.

\subsection{Auto-Parallels}

In Riemannian geometry, geodesics are paths that are local extrema of the metric distance (the shortest paths between points). This definition carries over to Weitzenb\"{o}ck space, with the geodesic equation being given by:
\begin{equation}
\ddot{x}^{\mu} + \chris{\mu}{\alpha}{\beta}\dot{x}^{\alpha}\dot{x}^{\beta} = 0,
\end{equation}

\noindent where $\chris{\mu}{\alpha}{\beta}$ are the Christoffel symbols of the metric.

\begin{figure}
\centering
\includegraphics[width=.8\textwidth]{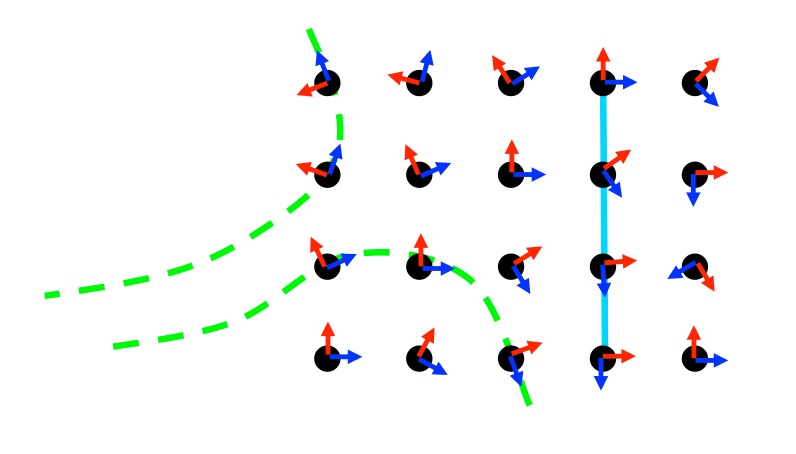}
\caption{The different between autoparallels (dashed lines) and geodesics (solid line) for a set of data where the local basis rotates (the basis is shown as two arrows at each point). The autoparallels follow the local basis.}
\label{fig:schematic}
\end{figure}

However, there are other curves that we can define in \weit space, which are curves that are as `straight' as possible, rather than as short as possible, which means following a constant direction with respect to the frame basis:
\begin{equation}
\dot{x}^{\mu}(t) = c^{a}e\indices{_{a}^{\mu}}(x(t)),
\end{equation}
where the $\{c^{a}\}$ are constants, and the curve has been parameterised to be constant speed. Figure~\ref{fig:schematic} shows the difference between geodesics and autoparallels for a simple two dimensional example.

If we now differentiate this a second time, and note that:
\[
\partial_{\nu}e\indices{_a^\mu} = -e\indices{_a^\beta}\Gamma\indices{^\mu_{\nu\beta}},
\]
then it is simple to show that this path is a solution of the auto-parallel equation:
\[
\ddot{x}^{\mu} + \Gamma\indices{^\mu_{\nu\beta}}\dot{x}^{\nu}\dot{x}^{\beta} = 0.
\]
This is an analogue of the geodesic equation with a Riemann-Cartan connection, rather than just the symmetric Levi-Civita connection.  In certain cases autoparallels and geodesics can coincide, but with the Euclidean metric this can only occur if the frames do not twist at all, which is when the torsion is identically zero.

Autoparallels in the \weit space map to straight lines in the Euclidean space. While in general the dimension of the Euclidean space will be higher (just as embedding a curved 2D surface in Euclidean space requires 3 (or more) dimensions, by the Nash Embedding Theorem) the dimensionality is the same for the flat \weit space. To construct the mapping, we take $\{x^{\mu}\}$ as Cartesian coordinates in $\mathbb{W}$ and use the vielbein to generate a map between tangent spaces of \weit space with torsion, and tangent spaces of Euclidean space, which are all equivalent. Then the mapping to Euclidean coordinates $\{y^{a}\}$ is given by:
\[
x^{\mu}(t) \:\: \mapsto \:\: y^{a}(t) = y^{a}(0) + \int\limits_{0}^{t}\dot{x}^{\mu}(s)e\indices{^a_\mu}(x(s))ds.
\]

Now two paths of the form:
\begin{eqnarray}
&&x(0) \rightarrow x(0) + \Delta x^{1} \rightarrow \left(x(0) + \Delta x^{1}\right) + \Delta x^{2}, \nonumber \\
&&x(0) \rightarrow x(0) + \Delta x^{2} \rightarrow \left(x(0) + \Delta x^{2}\right) + \Delta x^{1} \nonumber
\end{eqnarray}
end up at the same place in the \weit space, but not in the Euclidean space, where we have the path:
\begin{gather}
y^{a}(0) \rightarrow y^{a}(0) + e\indices{^a_1}(x(0))\Delta x^{1}
\rightarrow y^{a}(0) + e\indices{^a_1}(x(0))\Delta x^{1} +
e\indices{^a_2}(x(0)+\Delta x^{1})\Delta
x^{2} \nonumber \\
= y^{a}(0) + e\indices{^a_1}(x(0))\Delta x^{1} +
e\indices{^a_2}(x(0))\Delta x^{2} + \Delta
x^{1}\partial_{1}e\indices{^a_2}(x(0))\Delta x^{2}. \nonumber
\end{gather}
The difference between the paths is given by:
\[
\partial_{1}e\indices{^a_2} - \partial_{2}e\indices{^a_1} =
e\indices{^a_{\mu}}\left(\Gamma\indices{^{\mu}_{12}}-\Gamma\indices{^{\mu}_{21}}\right) = e\indices{^a_{\mu}}T\indices{^{\mu}_{12}},
\]
which is the mapping coefficients that have a non-vanishing curl, i.e., precisely the torsion. This construction is known as a non-holonomic mapping, for further details see \citep{Kleinert,Kleinert2,Guo}.

The construction we have given is based on the moving frame: a set of linearly independent vector fields  are derived and used to construct a derivative. However, for a dataset in an unknown manifold, this is not known a priori, rather, we need to build it from the data. Unfortunately, this is not possible. In the next section we describe an algorithm that iteratively generates this set in a way that is analagous to PCA, and give examples of its use on standard datasets. 

\section{PAPA: Principal AutoParallel Analysis}


The idea behind the algorithm is that autoparallels describe fibres (in the differential geometric sense of elements of the fibre bundle). The autoparallel gives a single direction along the vector field, which can be followed. It is therefore possible to choose a base space, and to identify the points where each fibre intersects that base space. This means that each datapoint can be projected onto the base space by constructing the autoparallel through it, and following until it intersects the base space, choosing the closest one if it meets it several times. 

The first direction has then been projected out, leaving data in a lower dimensional space. The same method can then be used to find the autoparallels in the new space, and the method can be iterated as many times as is required. A useful way to understand the algorithm is by analogy with the standard description (rather than implementation) of normal Principal Component Analysis (PCA): the first step is to find the direction with maximum variance as the first principal component, and then to successfully find components that maximise the remaining variance. 

PCA can be used for dimensionality reduction by retaining only a subset of the new components, with the data being projected into the subspace that they span. The choice of an appropriate number of components is often based on some percentage of the variance, or a similar consideration. For our method an appropriate choice is not so clear, but given that the noise is often isotropic, one appropriate time to stop is when no consistent autoparallel directions are found, but all directions seem to be equally preferable.

Note that  while at each stage we are approximating one vector field of the connection, it is not possible to construct the full connection from them, since the spaces that each vector field live in are different: they are not defined on the same space. 

Computationally, there are two tasks: to approximate the autoparallel, and to perform the projection onto the base space. 
We use standard PCA in a local neighbourhood to define a single direction along the vector field. This describes the local direction of the autoparallel, and we can therefore construct a numerical approximation to an autoparallel by iterating this approach: at a given point on the data manifold we compute the first local principal component, move a step of size $h$ in that direction, and then repeat. It is also possible to move in the negative direction, so that the entire data manifold can be traversed. 

\begin{figure}
\centering
\includegraphics[width=.4\textwidth]{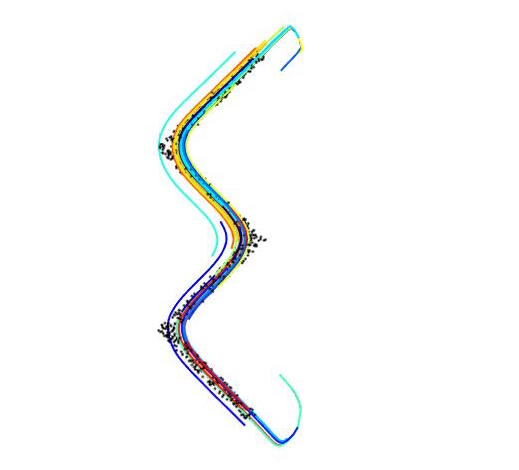}
\includegraphics[width=.4\textwidth]{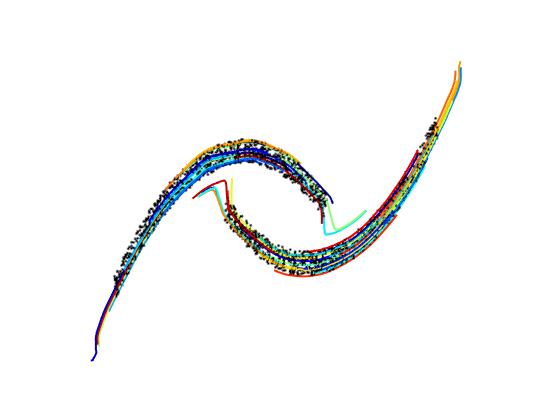}
\caption{Two dimensional examples of principal autoparallels. The underlying datasets consist of 1000 datapoints (shown as black dots). 50 autoparallels are then shown, starting from 50 randomly chosen datapoints, each autoparallel being followed forwards and backwards. The effects of the local PCA estimation at the endpoints can be seen in both cases: for the zigzags the curves overshoot, and then start to bend back towards the data, while for the boomerangs, on the end where the other part of the dataset appears, they head towards that data.}
\label{fig:zig}
\end{figure}

With any form of local PCA the size of the neighbourhood is critical, and that is particularly true here, since if too few datapoints are used the estimate of the direction of the autoparallel is noisy, while if too many datapoints are used the direction can be compromised by elements that are close by in Euclidean space, but not on the manifold; we will show an example of this shortly. We therefore estimate an appropriate size for the neighbourhood by computing the point-to-point distance (with a randomly datapoint selected as the origin). By plotting a histogram of these distances (an example is shown on the right of Figure~\ref{fig:swiss2}) it is possible to choose an appropriate value for the maximum distance allowable to be included in the neighbourhood.

\subsection{Demonstration}

As a first demonstration of our algorithm we use two standard two-dimensional datasets that are commonly used in manifold and surface learning. Figure~\ref{fig:zig} shows the datasets, each of which consists of 1000 datapoints drawn uniformly at random from the data distribution, together with a set of 50 autoparallels, each running for 200 steps ($h=0.01$) both forwards and backwards, and starting from randomly chosen datapoints.  Where there is good support in the data the autoparallels track the data well, but where the data stops, the autoparallels continue, as the local PCA allows overshooting, before the autoparallels bend round, back towards the dataset. In the dataset on the right, it can be seen that this effect of the local PCA means that the autoparallels can `fall off' the end of one curve and head towards the other portion of the data.

\begin{figure}
\centering
\includegraphics[width=.48\textwidth]{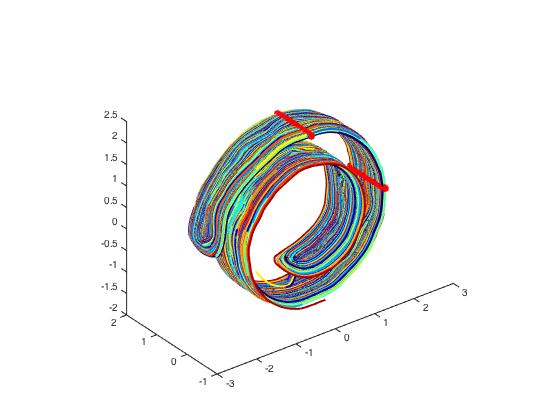}
\includegraphics[width=.48\textwidth]{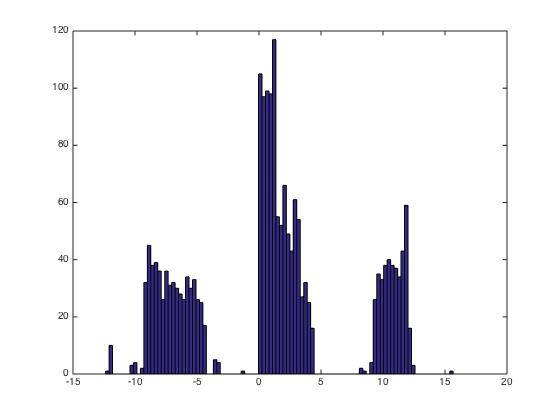}
\caption{{\em Left: }A set of 200 autoparallels on the swiss roll dataset. The points highlighted in red in the colour version (larger in greyscale) show the intersection of a plane running at $45^{\circ}$  to the $z$-axis. {\em Right: } A histogram of the signed distance of each datapoint from the nearest intersection with the plane. This would be the point that it would be projected onto.}
\label{fig:swiss1}
\end{figure}

The most famous dataset of manifold learning is the three-dimensional swiss roll, and we demonstrate that next. On the left of Figure~\ref{fig:swiss1} are drawn a set of 200 autoparallels following the data. It can be seen that they match the data well, and that there is sufficient data at the ends of the dataset for the autoparallels to curve around and run back along the dataset. Also visible in that figure are a set of points marking the intersections between a plane running at $45^{\circ}$  to the $z$-axis, which intersects with the dataset twice. Having identified these points it is possible to compute the signed distance from each datapoint to the nearest point of intersection, and the histogram of these distances are shown on the right of Figure~\ref{fig:swiss1}. The data can then be projected onto these two one-dimensional data spaces to complete the analysis. 

\begin{figure}
\centering
\includegraphics[width=.42\textwidth]{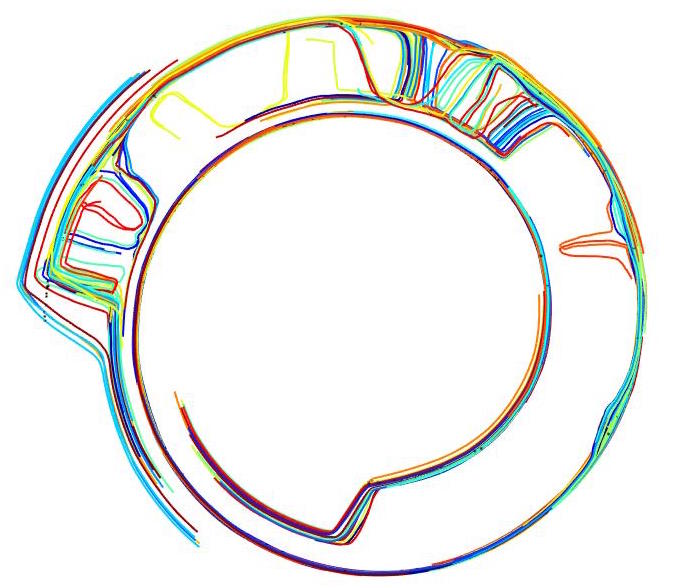}
\includegraphics[width=.42\textwidth]{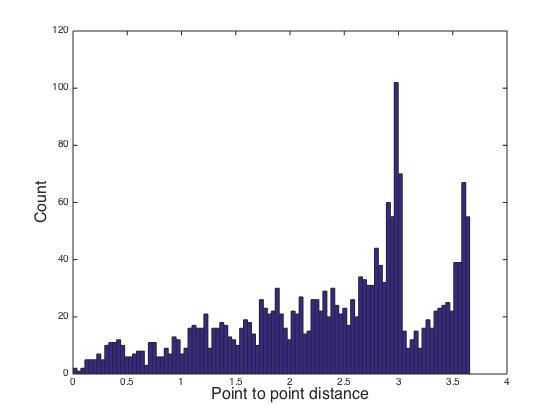}
\caption{{\em Left:} the effect of allowing the distance in the local PCA to get too large, with a side projection of the swiss roll dataset. At various points the random variation in the dataset means that a sufficiently large number of points from the next layer of the swiss roll are included, and so the approximate autoparallels wander off. {\em Right:} a histogram of the number of points with various nearest point-to-point distances, which can be used to set the maximum distance allowable for points in the local neighbourhood by looking for a point where there is a gap and then increase in the number of points found; based on this graph a value of 0.5 was chosen.}
\label{fig:swiss2}
\end{figure}

In order to demonstrate the difficulty of using local PCA for the construction of autoparallels, on the left of Figure~\ref{fig:swiss2} we demonstrate a case where the neighbourhood was set too large, meaning that datapoints from the next layer of the swiss roll were also selected, pulling the the autoparallel off the manifold. In order to choose an appropriate value, we computed the distribution of point-to-point distances to a random datapoint, and used this histogram to identify a suitable cut-off by looking for the first point where there was a clear drop-off in the number of points. This lead to the choice of a maximum of 0.5 being chosen in this case.
  
\begin{figure}
\centering
\includegraphics[width=.24\textwidth]{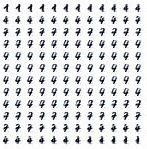}
\includegraphics[width=.24\textwidth]{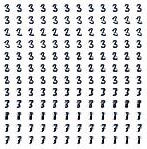}
\includegraphics[width=.24\textwidth]{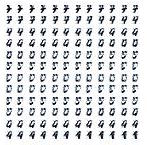}
\includegraphics[width=.24\textwidth]{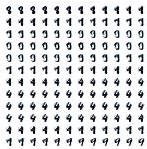}
\caption{Following autoparallels based on the digits data starting at randomly chosen instances of (left to right) 1, 3, 7, 8.  While the autoparallels do sometimes seem to form cycles through groups of digits, the progression along each autoparallel generally seems to describe a reasonable progression through the data.}
\label{fig:digits}
\end{figure}

As another demonstration of our algorithm, we used the \texttt{optdigits} dataset from the UCI Machine Learning Repository~\citep{UCI}. Figure~\ref{fig:digits} shows points along four autoparallels starting from randomly chosen examples of a 1, a 3, a 7, and an 8. The points along the autoparallels do seem to make some sense in terms of how they transition between the digits, although there are clearly some problems with the local PCA parameters, since the autoparallel starting at a 1 moves through 4 to 7, back to 4, and then back to 7, and the autoparallel starting at a 7 appears to do similar cycling between 0 and 5.   Overall, these results suggest that the autoparallels follow expected paths through the data.

\section{Related Algorithms} 

Conceptually, the algorithm that we find most similar to ours  is the Principal Curve algorithm, particularly the locally defined variant described in~\citet{Ozertem}. A principal curve is a curve through the data where each point on the curve is the mean of the data points that have it as the closest point on the curve~\citep{Hastie89}. 
In~\citet{Ozertem} they construct principal curves and surfaces based an estimate of the underlying continuum data density. This leads to an algorithm based on computing the eigenvectors of the Hessian (i.e., principal curvature directions where the curve is flat), with the constraint that the curve is orthogonal to the gradient, so that it sits on a ridge in the data. In order to project data onto the `nearest' principal curve they use the local covariance to identify the direction to travel. This local covariance is rather different to what we have considered in this paper. We are using the standard definitions of covariance and sample covariance, whereas in their paper they use the fact that for a Gaussian distribution they can find a combination of the Hessian and gradient that is not spatially varying, and that this can define a version of covariance. While the two versions are related, ours requires only a reasonable estimate of a single direction, whereas they need a sufficiently good density estimate to construct the local covariance matrix at each point.

Of course, our approach is also related to local PCA in its implementation. However, we are using locality to refer to points within the dataset, rather than treating the problem as a variant of $k$-means clustering. In local PCA it is common to approximate the whole data distribution by some set of PCA subspaces, and then has to decide which subspace to assign a point to (i.e., which cluster), then doing PCA again for each cluster~\citep{localPCA}. The method also differs from neighbourhood-preserving methods such as Local Linear Embedding~\citep{Roweis}, where locally flat patches of the manifold are identified and linked together, since we make no linear approximation. 

Another relevant dimensionality reduction algorithm is Independent Component Analysis (ICA)~\citep{Hyvarinen}, of which PAPA can be seen as a nonlinear variant, in the sense that sequential unwrapping of the global manifold identifies a series of nonlinear coordinate transforms. However, our approach has a strong ordering of the components, and does not require that the data is non-Gaussian. 

\section{Conclusion}

The analysis of data lying on differentiable manifolds provides many challenges not seen in the normal Euclidean case. Of particular importance is the local nature of all computations that are made, which is caused by the curvature of the space, meaning that parallel transport between points on the manifold are path dependent. In this paper we have introduced the idea of performing data analysis in \weit space, which is a differentiable manifold endowed with a connection that has torsion, but not curvature. The benefit of this space is that it is flat by construction; the price that is paid for this is that infinitesimal parallelograms do not close, the signature of torsion. 

Construction of the \weit space requires a position-dependent orthonormal frame to be defined on the manifold. We have described a constructive method to approximate autoparallels on the intrinsic data manifold by using the first local principal component at each datapoint, and then projecting the data into the subspace where that dimension is not included. We have provided experimental demonstrations that this algorithm works well on a variety of standard non-linear manifolds from the literature.

We will extend this approach to more complicated datasets, and also to the case of shape spaces, which has driven much of the creation of statistical methods for Riemannian manifolds.

\subsubsection*{Acknowledgments}

This work was supported by the Royal Society of New Zealand's Marsden Fund, and part of the work was done while Stephen Marsland was based at the Erwin Schr\"{o}dinger Institute for Mathematical Physics in Vienna, Austria.


\bibliographystyle{iclr2016_conference}
\bibliography{DataTorsion}

\end{document}